\input{aipcheck}

\documentclass[
    ,final            
  ]
  {aipproc}

\layoutstyle{6x9}

\begin{document}

\title{Tsallis distribution from minimally selected order statistics}



\author{G. Wilk}{
           address={The Andrzej So\l tan Institute for Nuclear Studies;
           Ho\.za 69; 00-681 Warsaw, Poland }
           }
\author{Z.W\l odarczyk}{
  address={Institute of Physics, \'Swi\c{e}tokrzyska Academy,
         \'Swi\c{e}tokrzyska 15; 25-406 Kielce, Poland and\\
         University of Arts and Sciences (WSU), Weso\l a 52, 25-353 Kielce,
         Poland}
         }

\begin{abstract}
We demonstrate that selection of the minimal value of ordered
variables leads in a natural way to its distribution being given
by the Tsallis distribution, the same as that resulting from
Tsallis nonextensive statistics. The possible application of this
result to the multiparticle production processes is indicated.
\end{abstract}

\maketitle

Tsallis distribution $p(E)$ of some variable $E$ is defined as
\begin{equation}
p_q(E) =
\frac{2-q}{T}\left[1-(1-q)\frac{E}{T}\right]^{\frac{1}{1-q}} ,
\label{Td}
\end{equation}
(with $E\in (0,\infty)$ for $q \ge 1$ and $E\in [0, T/(1-q)]$ for
$q <1$) where $T$ is some scale parameter (for example, if $E$ is
energy then $T$ is temperature). The mean value of variable $E$ is
$\langle E\rangle = T/(3-2q)$. In the limit $ q \rightarrow 1$
Tsallis distribution (\ref{Td}) becomes the usual exponential
(Boltzmann) distribution,
\begin{equation}
p_{q=1}(E) = \frac{1}{T} \exp \left( - \frac{E}{T} \right).
\label{ed}
\end{equation}

Distributions of type \eqref{Td} are ubiquitous in all fields of
research \cite{T-examples} \footnote{This includes also
applications to multiparticle production processes, which are of
special interest to us, see review \cite{qWilk}.}. Their origin is
rooted in the notion of nonextensive statistics introduced by
Tsallis \cite{T} (see also \cite{NEXT2005,T-examples} and
references therein), which for $q\rightarrow 1$ becomes the usual
Boltzmann-Gibbs one. There are numerous ways to obtain Tsallis
distribution \eqref{Td} which are discussed in the literature
\cite{NEXT2005} and from which we would like to mention here only
two. The first is based on the observation that some specific
intrinsic fluctuations of the parameter $T$ in the distribution
\eqref{ed} result in the eq. \eqref{Td}; in this case $(q-1)$
measures the strength of these fluctuations \cite{WW,BC}. This can
be contrasted with the second approach proposed in \cite{Kodama}
where suitable choice of particles in the phase space (introducing
effectively some specific correlations among them) also results in
eq. \eqref{Td} \footnote{Similar to this is the approach based on
the assumed fractality of phase space proposed in
\cite{fractality}.}. In this note we shall follow similar way of
reasoning and demonstrate that {\it distribution of the minimal
values} of some specific choices of variable $E$ (known in the
literature as {\it order statistics} \cite{OS}) also results in
Tsallis distribution \eqref{Td} but this time without necessity of
correlating the corresponding variables.\\

\medskip

Let us imagine therefore that we have to our disposal a number of
$n$ "ghost-particles" with energies $\varepsilon_{i=1,\dots,n}$,
with $\varepsilon_i$ following some distribution $f(\varepsilon)$.
Let us now order the values of $\varepsilon_i$, i.e., let us
introduce order statistics in this set. As result we are getting
the ordered set of energies, $\varepsilon_1 < \varepsilon_2 <
\dots < \varepsilon_n$, out of which we shall now choose the
actually observed ("real") particle defined as {\it particle with
the lowest energy}, $ E=\varepsilon_1 = \min (\{
\varepsilon_i\})$. Probability to find a particle with such energy
$E$ among $n$ elements is equal to $nf(\varepsilon)$ whereas
probability to find particle with energy exceeding $E$ is equal to
$1 - F(E)$, where $F(E) = \int_0^E d\varepsilon f(\varepsilon)$ is
distribuant of $f(\varepsilon)$. If particle of energy $E$ is
already that of the minimal energy it means that the remain $n-1$
particles have to posses higher energies. Probability of such
event is equal $[1 - F(E)]^{n-1}$. It means therefore that
distribution of the minimal value in the sample of $n$ elements is
\footnote{Notice that, because $f(E) = d F(E)$, distribution
$g(E)$ is properly normalized if $f(E)$ is normalized.}
\begin{equation}
g(E) = nf(E)[1- F(E)]^{n-1} . \label{g(E)}
\end{equation}
For some specific forms of distribution $f(\varepsilon$)
\eqref{g(E)} can be converted exactly into Tsallis distribution
\eqref{Td} with parameter $T$ being independent of $q$. We shall
in what follows discuss examples of such distributions, separately
for $q<1$ and $q>1$ cases.\\

For $q<1$ energies $\varepsilon$ of all $n$ particles considered
must be limited to the interval $0 < \varepsilon < (n-1)T$ ($T$ is
scale parameter mentioned before). Let us now assume that they are
distributed uniformly according to
\begin{equation}
f(\varepsilon) = \frac{1}{(n-1)T}.    \label{fqle1}
\end{equation}
Distribuant of this distribution is $F(E) = E/[(n-1)T]$, therefore
distribution of the minimal value \eqref{g(E)} takes the following
form
\begin{equation}
g(E) = \frac{n}{n-1}\frac{1}{T}\left( 1 -
\frac{1}{n-1}\frac{E}{T}\right)^{n-1} =
\frac{2-q}{T}\left[1-(1-q)\frac{E}{T}\right]^{\frac{1}{1-q}},
\label{g1le1}
\end{equation}
i.e., it coincides with Tsallis distribution with $q = (n-2)/(n-1)
< 1$ and $E<T/(1-q)$. Notice that increasing the range of allowed
$\varepsilon$, i.e., for $n \rightarrow \infty $, one obtains
exponential distribution \eqref{ed}.\\

For $q>1$ let us consider $n$ particles with unlimited energies
$\varepsilon$, $0 \leq \varepsilon < \infty$, assumed to be
distributed according to
\begin{equation} f(\varepsilon) =
\frac{c}{(1 + c\varepsilon)^2},\qquad c=\frac{1}{(n+1)T}.
\label{fqg1}
\end{equation}
Distribuant of this distribution is $F(E) = 1 - 1/(1+cE)$ and
distribution of the minimal value \eqref{g(E)} takes the form
\begin{equation}
g(E) = \frac{n}{n+1}\frac{1}{T}\left(1 +
\frac{1}{n+1}\frac{E}{T}\right)^{-(n+1)} =
\frac{2-q}{T}\left[1-(1-q)\frac{E}{T}\right]^{\frac{1}{1-q}},
\label{gqg1}
\end{equation}
i.e., again, the form of Tsallis distribution with $q =
(n+2)/(n+1) > 1$ and with no limit on $E$. Again, for
$n\rightarrow \infty$ one recovers the exponential distribution
\eqref{ed}.\\

\begin{figure}
  \includegraphics[height=.185\textheight]{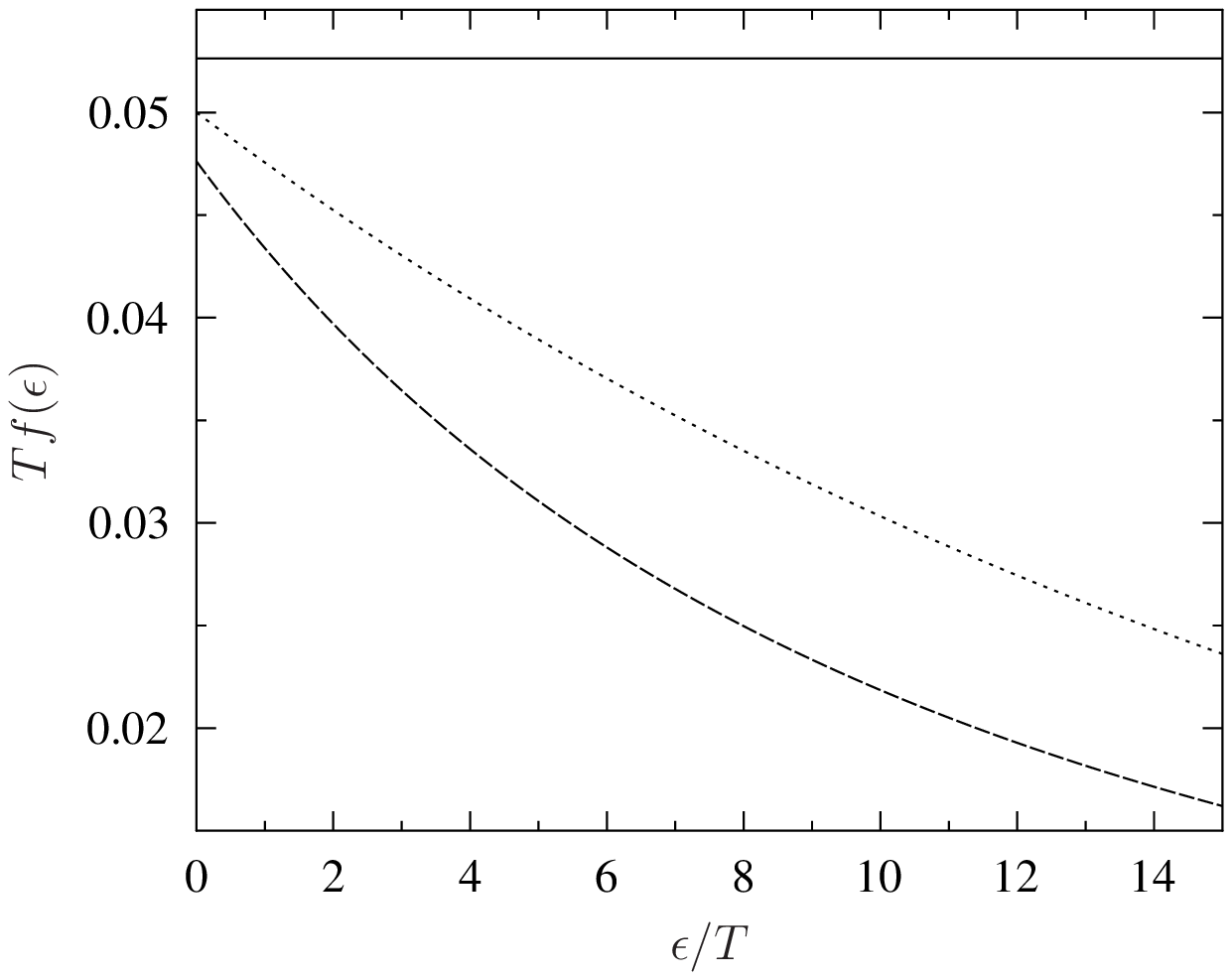}
  \includegraphics[height=.185\textheight]{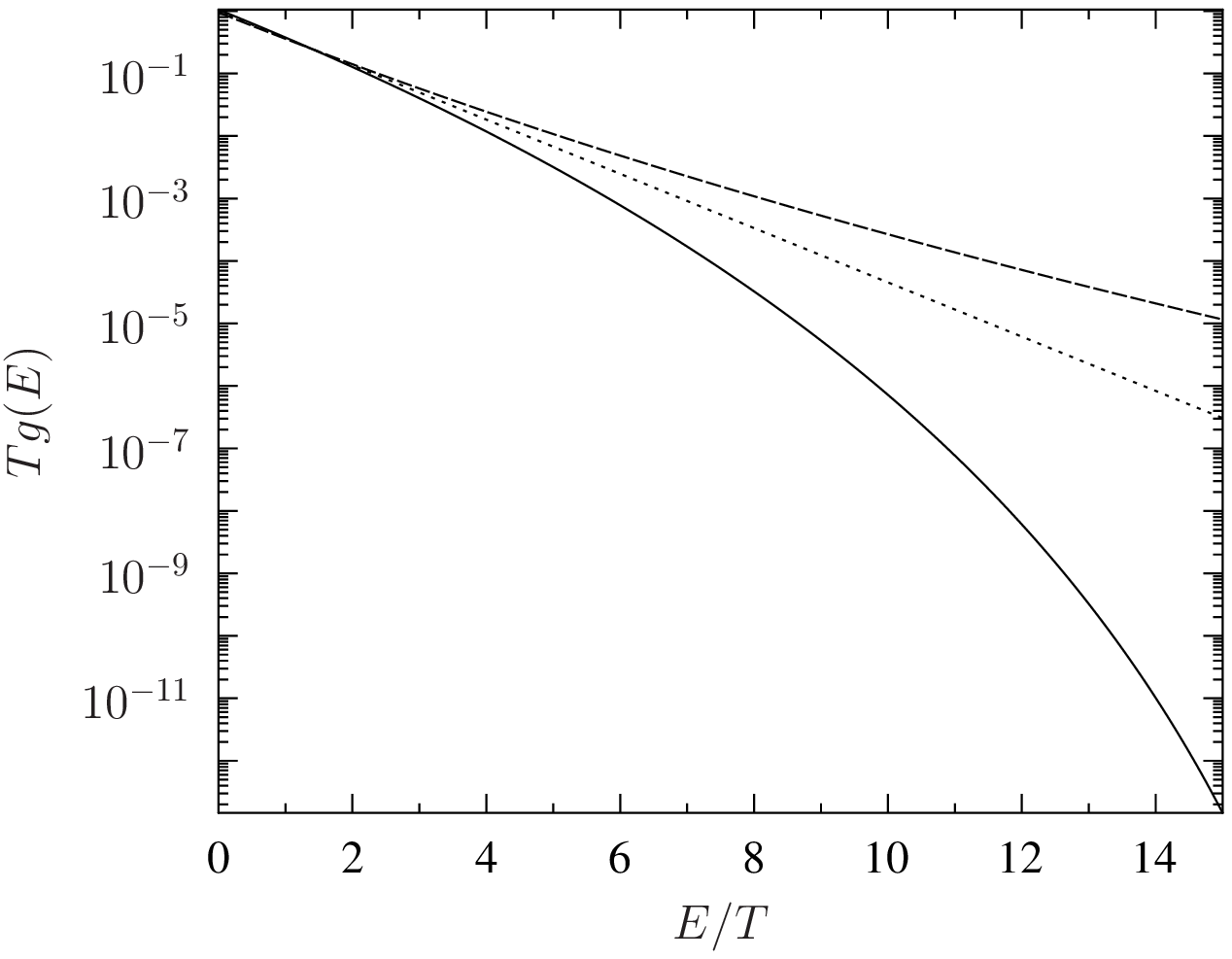}
  \includegraphics[height=.185\textheight]{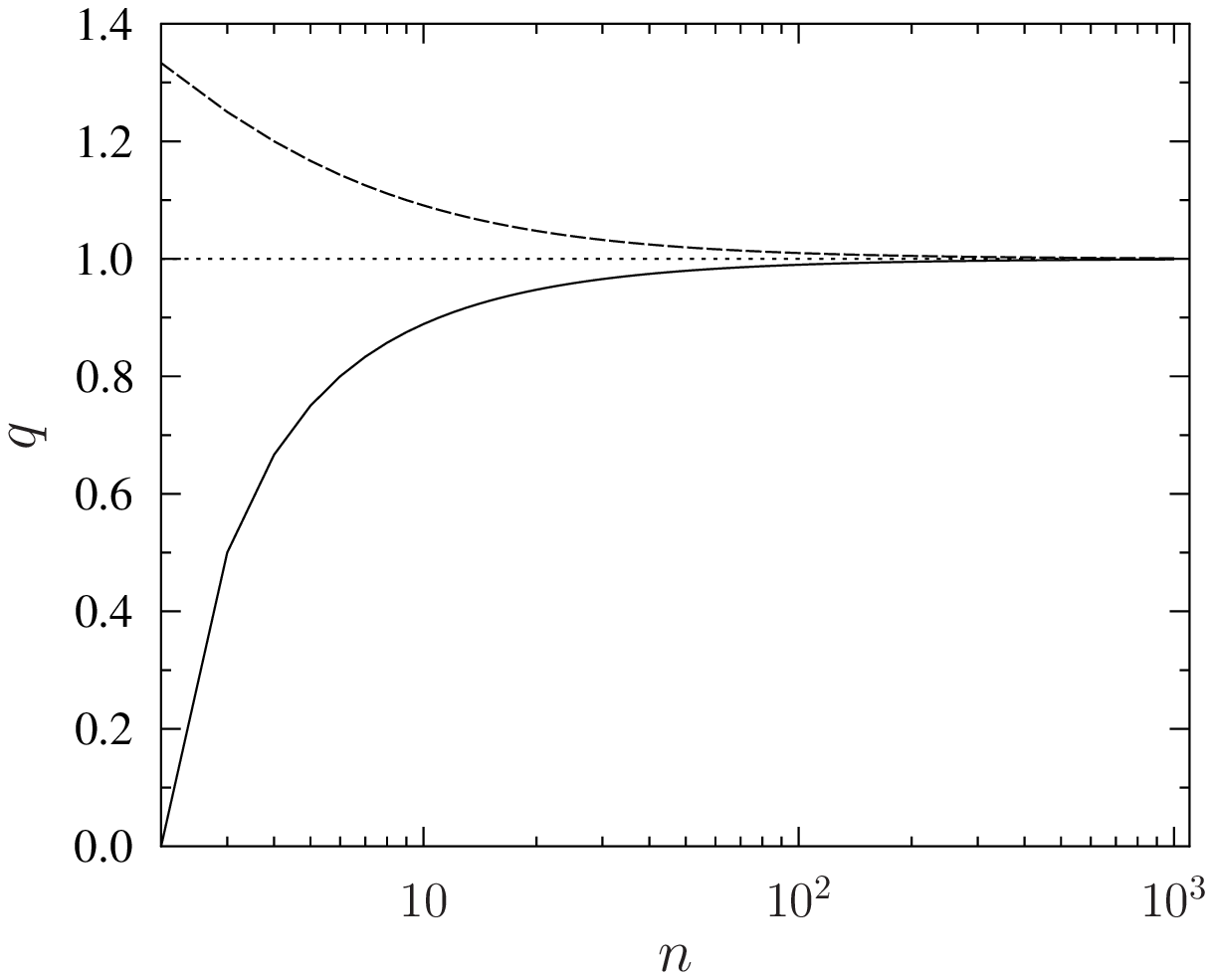}
  \caption{Examples of $T\, f(\varepsilon)$ versus $\varepsilon/T$ used here (left panel) and
  corresponding to them $T\, g(E)$ versus $E/T$ (middle panel) calculated for  $n = 20$.
  Continuous line is for $q < 1$, dotted line for $q = 1$ and dashed line for
  $q > 1$. The right panel displays the $q$ as function of $n$ corresponding to
  \eqref{g1le1} (continuous line) and \eqref{gqg1} (dashed line). They both converge
  to dotted line for $q=1$ for $n\rightarrow \infty$.}
  \label{fig:Fig1}
\end{figure}

Notice that in both cases for $n \rightarrow \infty$ one gets
distribution with $q=1$ (approached, respectively, from below or
from above). Actually, one can have the $q=1$ distribution
independently on the value of $n$ only for the exponential form of
the initial distribution of $\varepsilon$,
\begin{equation}
f(\varepsilon) = \frac{1}{nT} \exp\left( -
\frac{\varepsilon}{nT}\right). \label{fexp}
\end{equation}
In this case distribuant is $F(E) = 1 - \exp\left( -
E/(nT)\right)$ and distribution of the minimal value, $g(E)$, has
form of eq. \eqref{ed}, independent of $n$. The above results are
illustrated in Fig. \ref{fig:Fig1}.

\medskip

Let us now calculate the changes of entropy $S = - \int p(x)
\ln[p(x)] dx$ caused by the selection of the minimal value,
\begin{equation}
\Delta S_q = S_f -(<S'_f> + S_g ) .\label{DSq}
\end{equation}
Here $S_f$ is entropy of the initial distribution
$f(\varepsilon)$, which changes to $S'_f(E)$ after choosing the
minimal value of $\varepsilon$. It depends on $E$ because the
remaining values of $\varepsilon$ are all above $E$. Its average
over $g(E)$ is denoted by $<S'_f>$, whereas the entropy of the
selected particle is denoted by $S_g$. One has
\begin{equation}
\Delta S_q = \ln(n) - \frac{n}{n+1}\ln(T) -
\frac{n}{n+1}\ln(n+\xi) - \frac{(n-\xi)(n+2)}{(n+1)^2} +
\frac{1}{2}\xi (\xi -1)\left(\frac{1}{n} + \frac{1}{n+1}\right) ,
\label{sum}
\end{equation}
where $\xi = (-1,0,+1)$ for $q<1$, $q=1$ and $q>1$, respectively.
Notice that entropy always increases and that its increase is, in
the limit of $n\rightarrow \infty$, the same and equal $\Delta S_q
= -[1+\ln(T)]$. Because $S_f \neq \langle S'_f\rangle + S_g$ the
process of selection of the minimal value is nonextensive.\\

\medskip

To summarize: the order statistics is {\it per se} the important
branch of science \cite{OS} but we are in no position to discuss
it here. On the other hand, as shown in \cite{MGMG,qWilk}, the
selection of energies, very much alike to the one applied here,
when applied to hadronic physics results in the power-like
distributions observed in experiments. This fact has been
interpreted as indication of the need of some kind of new
statistical physics (Tsallis statistics) being at work here
\cite{qWW}. In fact, one can easily invent a non-thermal scenario
leading to thermal-like form of the observed spectra, see, for
example, recent work \cite{BH}. In such approach the resultant
distribution emerges not because of the equilibrating of energies
due to some collisions (i.e., because of the {\it kinematic
thermalization}) but rather because of the process of erasing of
memory of the initial state and is the result of the approaching
to a state of maximal entropy (called in \cite{BH} {\it stochastic
thermalization}). This seems to be very promising idea which needs
further investigations.

\begin{theacknowledgments}
Partial support (GW) of the Ministry of Science and Higher
Education under contracts 1P03B02230 and  CERN/88/2006 is
acknowledged.
\end{theacknowledgments}


\begin{thebibliography}{9}


\bibitem{T-examples} Cf., for example, C. Tsallis, \emph{Physica} \textbf{A340} 1--10 (2004),
                     \textbf{A344}, 718--736 (2004) and \textbf{A365}, 7--16 (2006) and references therein.
                     See also special issue of Europhysics News, Nov-Dec. 2005 (EPSP)
                     (http://www.europhysicsnews.com).

\bibitem{qWilk} G. Wilk, \emph{Braz. J. Phys.} \textbf{37} 714-716 (2007).

\bibitem{T} C.Tsallis, {\sl J. Stat. Phys.} {\bf 52} (1988) 479;
             cf. also C.Tsallis, {\sl Chaos, Solitons and Fractals}
             {\bf 13} (2002) 371, {\sl Physica} {\bf A305} (2002) 1
             and in {\it Nonextensive Statistical Mechanics and its
             Applications}, S.Abe and Y.Okamoto (Eds.), Lecture Notes
             in Physics LPN560, Springer (2000). For updated
             bibliography on this subject see
             http://tsallis.cat.cbpf.br/biblio.htm.

\bibitem{NEXT2005} Proc. of the NEXT-SigmaPhi 2005 Conference,
                   Kolymbari, Greece, 13-18 August 2005;
                   \emph{Physica} \textbf{A365} (1) (2006) and
                   \emph{Europ. Phys. J.} \textbf{B50} (1-2) (2006);
                   Editors: G. Kaniadakis, A. Carbone, M. Lissia.

\bibitem{WW} G.Wilk and Z.W\l odarczyk, \emph{Phys. Rev. Lett.}
             \textbf{84}, 2770--2773 (2000) and \emph{Chaos, Solitons and
             Fractals} \textbf{13/3}, 581--594 (2001).

\bibitem{BC} C. Beck and E.G.D. Cohen, \emph{Physica} \textbf{A322},
             267--275 (2003);  F. Sattin, \emph{Europ. Phys. J.} \textbf{B49},
             219--224 (2006).

\bibitem{Kodama} T. Kodama, \emph{J. Phys.} \textbf{G31}, S1051--S1056 (2005);
                 T.Kodama, H.-T.Elze, C.E.Augiar and T.Koide, \emph{Europ.
                 Lett.} \textbf{70}, 439--445 (2005).

\bibitem{fractality} V. Garcia-Morales and J. Pellicer,  \emph{Physica}
                     \textbf{A361}, 161--172 (2006).

\bibitem{OS} See, for example, H.A.David, \emph{Order statistics},
             Wiley-InterScience, 1981, or H.A.David and H.N.Nagaraja,
             \emph{Order Statistics}, (Wiley Series in
             Probability and Statistics),  Wiley-InterScience, 2003.

\bibitem{MGMG} M. Ga\'zdzicki and M.I. Gorenstein, \emph{Phys.
               Lett.} \textbf{B517} 250--254 (2001).

\bibitem{qWW} G.Wilk and Z.W\l odarczyk, \emph{Acta Phys. Polon.}
              \textbf{B35}, 2141-2146 (2004).

\bibitem{BH} P. Castorina, D. Kharzeev and H. Satz, {\it Thermal
             Hadronization and Hawking-Unruh Radiation in QCD},
             arXiv:0704.1426v1.




\end{thebibliography}
\end{document}